%% file: MS.tex
\documentclass[journal]{IEEEtran}

\usepackage[T1]{fontenc}
\usepackage{amsbsy}
\usepackage{amsmath}
\usepackage{fixmath}
\usepackage{amssymb}
\usepackage{siunitx}
\usepackage{graphicx}
\usepackage{subcaption}
\usepackage{booktabs}
\usepackage{cite}
\usepackage{cleveref}

\usepackage{algorithmic}
\usepackage{algorithm}
\usepackage{tabularx}
\usepackage{color}

\newlength\myindent
\usepackage{tikz}
\usetikzlibrary{arrows,decorations,backgrounds,shadows,plotmarks,calc}

\usepackage{pgfplots}
\pgfplotsset{compat=newest}
\pgfplotsset{plot coordinates/math parser=false}
\pgfplotsset{every axis/.append style={font=\footnotesize}}
\newlength\figureheight
\newlength\figurewidth
\DeclareMathOperator*{\h}{H}
\DeclareMathOperator*{\T}{T}
\DeclareMathOperator*{\argmin}{argmin}
\DeclareMathOperator*{\argmax}{argmax}

\DeclareMathOperator{\Quant}{Q}
\DeclareMathOperator{\E}{E}

\DeclareMathOperator{\sign}{sign}

\DeclareMathOperator{\vectorize}{vec}
\DeclareMathOperator{\normcdf}{\Phi}

\newcommand{\Tr}{\operatorname{T}}
\DeclareMathOperator*{\diag}{diag}
\DeclareMathAlphabet{\mathbit}{OML}{cmr}{bx}{it}
\DeclareMathAlphabet{\mathsf}{OT1}{cmss}{m}{n}
\DeclareMathAlphabet{\mathbsf}{OT1}{cmss}{bx}{it}

\definecolor{ppink}{HTML}{c71829}%
\definecolor{bblue}{HTML}{0496ff}%
\definecolor{bblack}{HTML}{2d3047}%
\definecolor{oorange}{HTML}{ffbc42}%

\begin{document}
\title{Efficient Non-linear Equalization$\;$for 1-bit Quantized Cyclic$\;$Prefix-Free Massive MIMO Systems}
\author{Daniel~Plabst$^1$, Jawad~Munir$^1$, Amine Mezghani$^2$ and Josef~A.~Nossek$^{1,3}$\\ %
$^1$Associate Professorship of Signal Processing, Technische Universit\"at M\"unchen, 80290 Munich, Germany\\
$^2$Wireless Networking and Communications Group, The University of Texas at Austin, Austin, TX 78712, USA\\
$^3$Department of Teleinformatics Engineering, Federal University of Cear\'{a}, Fortaleza, Brazil\\
Email: \{daniel.plabst, jawad.munir, josef.a.nossek\}@tum.de, amine.mezghani@utexas.edu
}

\maketitle

\begin{abstract}

This paper addresses the problem of data detection for a massive \textit{Multiple-Input-Multiple-Output} (MIMO) base station which utilizes 1-bit \textit{Analog-to-Digital Converters} (ADCs) for quantizing the uplink signal. The existing literature on quantized massive MIMO systems deals with \textit{Cyclic Prefix} (CP) transmission over frequency-selective channels. In this paper, we propose a computationally efficient block processing equalizer based on the \textit{Expectation Maximization} (EM) algorithm in CP-free transmission for 1-bit quantized systems. We investigate the optimal block length and overlapping factor in relation to the \textit{Channel Impulse Response} (CIR) length based on the \textit{Bit Error-Rate} (BER) performance metric.

As EM is a non-linear algorithm, the optimal estimate is found iteratively depending on the initial starting point of the algorithm. Through numerical simulations we show that initializing the EM-algorithm with a Wiener-Filter (WF) estimate, which takes the underlying quantization into account, achieves superior BER-performance compared to initialization with other starting points.

\end{abstract}

\IEEEpeerreviewmaketitle

\input{Introduction}
\input{Channel_Model}
\input{EM_Nonlinear_Data_Equalization}
\input{Computational_Complexity}
\input{Simulation_Results_And_Analysis}
\input{Conclusion}
%
\appendices
\input{Derivation_of_E_and_M_Step}
\vspace*{-0.50cm}

\ifCLASSOPTIONcaptionsoff
  \newpage
\fi

\bibliographystyle{IEEEtran}
\vspace{10pt}
\bibliography{MS}

\end{document}

%% file: Introduction.tex
\section{Introduction}\label{sec1}

Massive MIMO plays an important role for future communication systems, since the large number of antennas is capable of increasing the spectral efficiency and the amount of useable spectrum \cite{Larsson2014}.
However, a simple and power-efficient analog \textit{Radio Frequency}
(RF)-frontend design, together with the use of appropriate
baseband-processing algorithms, becomes crucial to support a large
number of antennas.
Especially in high-speed millimeter-wave
communication systems, the increasing complexity and power-consumption
of the key components in the RF-chain, such as the high-speed ADC, can
be identified as the primary bottlenecks. Whereas the
power-consumption of the ADC scales roughly exponentially with the
number of quantization bits \cite{Murmann2016}, the use of $1$-bit ADCs consumes the least amount of power and simplifies the hardware-complexity of the entire RF-frontend significantly.
The lost information due to the coarse quantization can furthermore be recovered by designing data-detection algorithms, which take the effect of coarse quantization into consideration.

Equalization algorithms for narrowband systems with frequency-flat channels have been investigated in \cite{Choi2016} and \cite{Risi2014} for the case of $1$-bit ADCs at the receive antennas.
In \cite{Studer2016}, C. Struder and G. Durisi have recently proposed algorithms for quantized  maximum  a-posteriori  (MAP)  channel estimation and  data detection under frequency-selective channels. In \cite{C_Mollen_2016}, low-complexity channel estimation and data detection for frequency-selective massive MIMO systems employing $1$-bit ADCs was proposed based on linear combiners. A message passing algorithm for data-detection for an underlying quantized single-carrier system is proposed in \cite{7037311}. To the best of our knowledge, all the mentioned contributions in the massive MIMO literature utilize CP for \textit{Orthogonal Frequency Division Multiplexing} (OFDM) and \textit{Single Carrier} (SC) transmission techniques, i.e., CP-OFDM and CP-SC.

State-of-the-art communication systems apply CP for efficient \textit{Frequency Domain Equalization} (FDE) by means of the FFT. However, the use of a CP comes at the price of a loss in spectral efficiency. It is therefore desirable to investigate computationally-efficient equalization methods without CP.
In \cite{Munir_Plabst18} we have proposed efficient linear-FDE for $1$-bit quantized wide-band massive MIMO systems without CP,  using an overlap-save method for equalization, while taking the quantization effect into consideration.

The EM-algorithm was originally introduced for channel estimation in wide-band \textit{Single-Input-Single-Output} SISO systems \cite{LOK_EM} and extended to flat-fading channels for massive MIMO systems in \cite{Mezghani2010} and \cite{ivrlac2007mimo}. In \cite{Mo_Schniter_Heath_EM} the authors propose a slightly modified EM-algorithm for mmWave frequency-flat MIMO channels, which is found to have a high computational complexity based on a matrix inversion in time-domain \cite{Mo_Schniter_Heath_EM_VAMP}.
In \cite{Stoeckle_EM} channel estimation based on the EM-algorithm is performed for frequency-selective massive MIMO systems, however, under the reassumption of CP.
In this work, we therefore propose efficient nonlinear-FDE using the EM-algorithm for $1$-bit quantized, frequency-selective, massive MIMO systems without CP.

The paper is organized as follows: An exact and mismatched quantized system model is introduced in Section~\ref{sec2}. Section~\ref{sec3} describes a probabilistic model for data-detection. In Section~\ref{sec4} and~\ref{sec5} we derive the time-domain and frequency-domain representation of the EM-algorithm, respectively, and assess their complexity. In Section~\ref{sec6}, the performance of EM is evaluated and compared to linear equalization methods. Section~\ref{sec7} concludes the paper.

Notation: Bold letters indicate vectors and matrices, non-bold letters express scalars. For a matrix $\boldsymbol{A}$, we denote complex conjugate, transpose and Hermitian transpose by
$\boldsymbol{A}^*$, $\boldsymbol{A}^{\T}$ and $\boldsymbol{A}^{\h}$, respectively. The operator  $\diag\left(\boldsymbol{A}\right)$ describes a diagonal matrix containing only the diagonal elements of $\boldsymbol{A}$ and
$\text{vec}{\left(\boldsymbol{A}\right)}$ denotes the vectorization operation with column-major order. The Kronecker product between matrices is given as $\boldsymbol{A}\otimes \boldsymbol{B}$.
The $n \times n$ identity matrix is denoted by $\boldsymbol{\text{I}}_{n}$, while the $n \times m$ all-zeros matrix is defined as $\boldsymbol{0}_{n \times m}$.

%% file: Channel_Model.tex
\section{System Model}\label{sec2}

The uplink of a single-cell scenario is considered where the \textit{Base-Station}  (BS) equipped with $M$ antennas receives the signals from $K$ single-antenna \textit{Mobile-Stations} (MSs). We assume a frequency-selective block fading channel between each pair of MS and BS antennas. The channel between BS $m\in\left\{1,2,\ldots,M\right\}$ and MS $k\in\left\{1,2,\ldots,K\right\}$ is completely characterized by an impulse response of $L+1$ taps, denoted by $\boldsymbol{h}_{mk} \in\mathbb{C}^{\left(L+1\right)\times 1}$.

We will derive the input-output relationship based on the exact and the mismatched model in the next two subsections.
\subsection{Exact Model (ExaMod): Block-Toeplitz Channel Matrix}
The unquantized receive signal at BS $m$ is written as:
\begin{align}
y_{m}\left[n\right] %
&= \sum_{l=0}^{L}{\boldsymbol{h}_{m}^{\T}[l]\boldsymbol{x}\left[n-l\right]}+\eta_m\left[n\right],
\label{eq:input-output-MSs-BS_antenna_data_eq}
\end{align}
where
$\boldsymbol{x}\left[n\right]=\begingroup %
\setlength\arraycolsep{4pt}
\begin{bmatrix}
x_{1}\left[n\right] & x_{2}\left[n\right] & \cdots & x_{K}\left[n\right]
\end{bmatrix}^{\Tr}
\endgroup \!\in\mathbb{C}^{K\times 1}$
is the zero-mean circularly-symmetric complex valued transmit vector with
$\mathrm{E}_{\boldsymbol{x}}\left[\boldsymbol{x}\left[n\right]\cdot\boldsymbol{x}^{\h}\left[n\right]\right]{}\!=\!\sigma^{2}_{x}\boldsymbol{I}_K$
and
$\boldsymbol{h}_{m}\left[l\right]=\begin{bmatrix}{h}_{m1}\left[l\right] & {h}_{m2}\left[l\right] & \cdots & {h}_{mK}\left[l\right]\end{bmatrix}^{\Tr}\!\!\in\mathbb{C}^{K\times 1}$
is constructed from the $l^{\text{th}}$ tap of the channel impulse response from all users on the $m^{\text{th}}$ antenna.
Let the noise be drawn from the i.i.d. zero-mean circularly-symmetric complex Gaussian vector %
$\boldsymbol{\eta}\left[n\right]=\begin{bmatrix}\eta_{1}\left[n\right] & \eta_{2}\left[n\right] & \cdots & \eta_{M}\left[n\right]\end{bmatrix}^{\Tr}\in\mathbb{C}^{M\times 1}$,
having the noise-covariance of $\mathrm{E}_{\boldsymbol{\eta}}\left[\boldsymbol{\eta}\left[n\right]\cdot\boldsymbol{\eta}\left[n\right]^{\h}\right]{}=\sigma^{2}_{\eta}\boldsymbol{I}_M$.
We furthermore assume that the transmit and noise symbols are temporarily uncorrelated.

Using \eqref{eq:input-output-MSs-BS_antenna_data_eq}, the unquantized receive vector
$\boldsymbol{y}\left[n\right]=\begin{bmatrix}y_{1}\left[n\right] & y_{2}\left[n\right] & \cdots & y_{M}\left[n\right]\end{bmatrix}^{\Tr}\in\mathbb{C}^{M\times 1}$
at time instant $n$ can be written as
\begin{equation}
\boldsymbol{y}[n]=\sum_{l=0}^{L}\boldsymbol{H}_l\boldsymbol{x}[n-l]+\boldsymbol{\eta}[n],\label{eq:compact_spatial_MIMO_rec_time_n}
\end{equation}
where
$
\boldsymbol{H}_l=\begin{bmatrix}\boldsymbol{h}_{1}\left[l\right]& \boldsymbol{h}_{2}\left[l\right] & \cdots & \boldsymbol{h}_{M}\left[l\right]\end{bmatrix}^{\Tr}\in\mathbb{C}^{M\times K}
$
is a channel impulse response matrix. The signal vector $\boldsymbol{y}[n]$ is then quantized by a $1$-bit \emph{uniform scalar} quantizer to obtain
\begin{align}
\boldsymbol{r}[n] &=
\Quant\left(\boldsymbol{y}[n]\right)=\Quant\left(\sum_{l=0}^{L}\boldsymbol{H}_l\boldsymbol{x}[n-l]+\boldsymbol{\eta}[n]\right),
\label{eq:1_bit_quantized_compact_spatial_MIMO_rec_time_n}
\end{align}
where $\Quant\left(\cdot\right)$ is applied element-wise to $\boldsymbol{y}[n]$ and keeps only the sign of the real and imaginary part, i.e., $\Quant\left(z\right) = \sign\left(\Re\left\{z\right\}\right)+\textrm{j}\sign\left(\Im\left\{z\right\}\right)$ for $z\in\mathbb{C}$ with
\begin{equation*}
\sign:~\mathbb{R}\to\left\{-1,+1\right\},
x\mapsto\sign\left(x\right) =
\begin{cases}
-1,&x\leq0\\
+1,&x>0
\end{cases}.
\end{equation*}
Let us collect $N_b$ vectors, with a condition that $N_b > L$, corresponding to time instances $n,n-1,\ldots,n-\left(N_b-1\right)$ in order to form a space-time quantized receive matrix $\boldsymbol{R}[n]$, unquantized receive matrix $\boldsymbol{Y}[n]$, and noise matrix $\boldsymbol{N}[n]$ as:
\begin{align*}
\boldsymbol{R}[n]  &=
\begingroup %
\setlength\arraycolsep{4pt}
\begin{bmatrix}
\boldsymbol{r}\left[n\right] & \boldsymbol{r}\left[n-1\right] &\cdots & \boldsymbol{r}\left[n-\left(N_b-1\right)\right]
\end{bmatrix}\in\mathbb{C}^{M \times N_b},
\endgroup\\
\boldsymbol{Y}[n]  &=
\begingroup %
\setlength\arraycolsep{4pt}
\begin{bmatrix}
\boldsymbol{y}\left[n\right] & \boldsymbol{y}\left[n-1\right] &\cdots & \boldsymbol{y}\left[n-\left(N_b-1\right)\right]
\end{bmatrix}\in\mathbb{C}^{M \times N_b},
\endgroup\\
\boldsymbol{N}[n]&=
\begingroup %
\setlength\arraycolsep{4pt}
\begin{bmatrix}
\boldsymbol{\eta}\left[n\right] & \boldsymbol{\eta}\left[n-1\right] &\cdots & \boldsymbol{\eta}\left[n-\left(N_b-1\right)\right]
\end{bmatrix}\in\mathbb{C}^{M \times N_b}.
\endgroup
\end{align*}
The transmit matrix  $\boldsymbol{X}[n]\in\mathbb{C}^{K\times \left(N_b+L\right)}$ is given by %
\begin{align}
\boldsymbol{X}[n]&=
\begin{bmatrix}
\boldsymbol{X}_{\textrm{c}}\left[n\right] & \color{black}{\boldsymbol{X}_{\textrm{in}}\left[n\right]}
\end{bmatrix}, \quad \textrm{where}\\
\boldsymbol{X}_{\textrm{c}}[n]&\!=\!
\begingroup %
\setlength\arraycolsep{2.5pt}
\begin{bmatrix}
\boldsymbol{x}\left[n\right] & \boldsymbol{x}\left[n\!-\!1\right]\!&\!\cdots\!&\! \boldsymbol{x}\left[n\!-\!\left(N_b\!-\!1\right)\right]
\end{bmatrix}
\endgroup \in\mathbb{C}^{K\times N_b}\!,\!\label{eq:KNb_transmit_vector}\\
\boldsymbol{X}_{\textrm{in}}[n]&\!=\!
\begingroup %
\setlength\arraycolsep{2.5pt}
\begin{bmatrix}
\boldsymbol{x}\left[n\!-\!N_b\right] & \cdots & \boldsymbol{x}\left[n\!-\!\left(N_b\!-\!1\!+\!L\right)\right]
\end{bmatrix}
\endgroup\in\mathbb{C}^{K\times L},\label{eq:KL_transmit_vector}
\end{align}
such that the space-time input-output relationship of the quantized MIMO system is given as
\begin{align}
\vectorize\{\boldsymbol{Y}[n]\}&= \boldsymbol{\check{H}}\vectorize\{\boldsymbol{X}[n]\}+\vectorize\{\boldsymbol{N}[n]\}\in\mathbb{C}^{M\cdot N_b\times 1}, \label{eq:vectorization_unquantized_MIMO_sc_data_equalization_model} \\
\vectorize\{\boldsymbol{R}[n]\}&=\Quant\left(\boldsymbol{\check{H}}\vectorize\{\boldsymbol{X}[n]\}+\vectorize\{\boldsymbol{N}[n]\}\right)\label{eq:quantized_MIMO_sc_data_equalization_model},
\end{align}
where the channel matrix $\boldsymbol{\check{H}}\in\mathbb{C}^{M\cdot N_b\times K\left(N_b+L\right)}$ has a block-Toeplitz structure of the form
\begin{equation}
\boldsymbol{\check{H}}=
\begingroup %
\setlength\arraycolsep{2pt}
\begin{bmatrix}
\boldsymbol{H}_{0} & \boldsymbol{H}_{1} & \cdots & \boldsymbol{H}_{L} & \boldsymbol{0}& \cdots& & \color{black}{\ldots}&\color{black}{\boldsymbol{0}}  \\
\boldsymbol{0} & \ddots 			& 			 & 										& \ddots 				&		\ddots		& & &\color{black}{	\vdots}\\
 &		\ddots					& \boldsymbol{H}_{0} & \ldots	& \ldots 				& \boldsymbol{H}_{L}&\color{black}{\boldsymbol{0}} & & \\
 & 					  	& 							 	 &\boldsymbol{H}_{0} &\ldots & \boldsymbol{H}_{L-1}& \color{black}{\boldsymbol{H}_{L}}& &\\
\vdots				&			&											&  									& \ddots			& \vdots & & \ddots&\\
\boldsymbol{0}	& \ldots		&  &	&	\boldsymbol{0}	& \boldsymbol{H}_{0}&\color{black}{\boldsymbol{H}_{1}}&\ldots &\color{black}{\boldsymbol{H}_{L}}
\label{eq:SC_toeplitz_channel_matrix}
\end{bmatrix}.
\endgroup
\end{equation}
Here, the matrix $\boldsymbol{0}$ denotes  $\boldsymbol{0}_{M\times K }$ for the sake of brevity.
\subsection{Mismatched Model (MisMod): Block-Circulant Channel Matrix Approximation}
The first step in the mismatched model is to represent a \emph{block-Toeplitz} channel matrix in the system model \eqref{eq:vectorization_unquantized_MIMO_sc_data_equalization_model} as a \emph{block-circulant} channel matrix with an interference term:
\begin{align}
\vectorize\{\boldsymbol{Y}[n]\}
&=\boldsymbol{\check{H}}_{\textrm{cir}}\vectorize\{\boldsymbol{X}_{\textrm{c}}[n]\}+\vectorize\{\boldsymbol{N}[n]\}+\boldsymbol{\check{\gamma}}'[n]. \label{eq:vectorization_unquantized_MIMO_sc_data_equalization_model_block_circulant}
\end{align}
In \eqref{eq:vectorization_unquantized_MIMO_sc_data_equalization_model_block_circulant}, $\boldsymbol{\check{H}}_{\textrm{cir}}\in\mathbb{C}^{M\cdot N_b \times K\cdot N_b}$ is a \emph{block-circulant} matrix
\begin{equation}
\boldsymbol{\check{H}}_{\textrm{cir}}=
\begingroup %
\setlength\arraycolsep{2pt}
\begin{bmatrix}
\boldsymbol{H}_{0} & \boldsymbol{H}_{1} & \cdots & \boldsymbol{H}_{L} & \boldsymbol{0}& \cdots  \\
									 & \ddots 						& 			 & 										& \ddots 				&					\\
								& 									& \boldsymbol{H}_{0} & \ldots	& \ldots 				& \boldsymbol{H}_{L} \\
\color{black}{\boldsymbol{H}_{L}} &									& 							 	 &\boldsymbol{H}_{0} &\ldots & \boldsymbol{H}_{L-1}\\
\vdots							&	\ddots					&											&  									& \ddots			& \vdots \\
\color{black}{\boldsymbol{H}_{1}}	& \ldots 					& \color{black}{\boldsymbol{H}_{L}} & 										&					& \boldsymbol{H}_{0}
\end{bmatrix}
\endgroup \textrm{ and},
\end{equation}
$\boldsymbol{\check{\gamma}}'[n]=\boldsymbol{\check{H}}'_{\textrm{in}}(\begin{bmatrix}{\vectorize\{\boldsymbol{X}_{\textrm{in}}[n]\}}^{\Tr}\,\,\,\,\boldsymbol{0}_{1\times \left(N_b-L\right)K}
\end{bmatrix}^{\Tr}\!-\vectorize\{\boldsymbol{X}_{\textrm{c}}[n]\})$ can be considered as an interference noise which corrupts the last  $M\cdot L$ equations in \eqref{eq:vectorization_unquantized_MIMO_sc_data_equalization_model_block_circulant}, where $\boldsymbol{\check{H}}'_{\textrm{in}}\in\mathbb{C}^{M\cdot N_b \times K\cdot N_b}$ is given as:
\begin{equation}
\boldsymbol{\check{H}}'_{\textrm{in}}=
\begingroup %
\setlength\arraycolsep{2pt}
\begin{bmatrix}
\boldsymbol{0} & \boldsymbol{0} & \cdots & \boldsymbol{0} & \boldsymbol{0}& \cdots  \\
									 & \ddots 						& 			 & 										& \ddots 				&					\\
								& 									& \boldsymbol{0} & \ldots	& \ldots 				& \boldsymbol{0} \\			\color{black}{\boldsymbol{H}_{L}} &								& 							\boldsymbol{0}& &\ldots & \boldsymbol{0}\\
\vdots							&	\ddots					&											&  									& \ddots			& \vdots \\
\color{black}{\boldsymbol{H}_{1}}	& \ldots 					& \color{black}{\boldsymbol{H}_{L}} & 										&					& \boldsymbol{0}
\end{bmatrix}.
\endgroup
\end{equation}
We can now obtain a mismatched model by ignoring the interference term, i.e.,
\begin{align}
\vectorize\{\boldsymbol{Y}[n]\}&\approx \boldsymbol{\check{H}}_{\textrm{cir}}\vectorize\{\boldsymbol{X}_{\textrm{c}}[n]\}+\vectorize\{\boldsymbol{N}[n]\}, \label{eq:vectorization_unquantized_MIMO_sc_data_equalization_model_block_circulant_final} \\
\vectorize\{\boldsymbol{R}[n]\}&\approx \Quant\left(\boldsymbol{\check{H}}_{\textrm{cir}}\vectorize\{\boldsymbol{X}_{\textrm{c}}[n]\}+\vectorize\{\boldsymbol{N}[n]\}\right). \label{eq:vectorization_quantized_MIMO_sc_data_equalization_model_block_circulant_final}
\end{align}
In Section~\ref{sec4} we will show that using the mismatched model \eqref{eq:vectorization_quantized_MIMO_sc_data_equalization_model_block_circulant_final} will enable a computationally efficient inversion of $\boldsymbol{\check{H}}_{\textrm{cir}}$ in the frequency domain.

\section{Probabilistic Model for Data Detection}
\label{sec3}
This section derives the joint \textit{Probability Density Function} (PDF) between the transmit signal $\boldsymbol{X}[n]$, unquantized receive signal $\boldsymbol{Y}[n]$ and quantized receive signal $\boldsymbol{R}[n]$. Let us generically express the exact (ExaMod) and the mismatched (MisMod) system model as follows:
\begin{align}
&\vectorize\{\boldsymbol{Y}[n]\}=\boldsymbol{A}\vectorize\{\boldsymbol{\chi}[n]\}+\vectorize\{\boldsymbol{N}[n]\},\label{eq:general_vectorization_quantized_MIMO_sc_data_equalization_model_block}\\
&\vectorize\{\boldsymbol{R}[n]\}=\Quant\left(\vectorize\{\boldsymbol{Y}[n]\}\right),
\end{align}
where $\boldsymbol{A}=\boldsymbol{\check{H}}\in\mathbb{C}^{M\cdot N_b\times K\cdot (N_b+L)}$, $\boldsymbol{\chi}[n]=\boldsymbol{X}[n]\in\mathbb{C}^{K\times (N_b+L)}$ for the exact system model \eqref{eq:vectorization_unquantized_MIMO_sc_data_equalization_model}, and $\boldsymbol{A}=\boldsymbol{\check{H}}_{\textrm{cir}}\in\mathbb{C}^{M\cdot N_b\times K\cdot N_b}$, $\boldsymbol{\chi}[n]=\boldsymbol{X}_{\textrm{c}}[n]\in\mathbb{C}^{K\times N_b}$ for the mismatched system model \eqref{eq:vectorization_unquantized_MIMO_sc_data_equalization_model_block_circulant_final}.

For the sake of brevity, let us represent: $\boldsymbol{\check{y}}=\vectorize\{\boldsymbol{Y}[n]\}$, $\boldsymbol{\check{\xi}}=\vectorize\{\boldsymbol{\chi}[n]\}$, $\boldsymbol{\check{\eta}}=\vectorize\{\boldsymbol{N}[n]\}$, $\boldsymbol{\check{r}}=\vectorize\{\boldsymbol{R}[n]\}$ and $\boldsymbol{\check{z}}=\boldsymbol{A}\boldsymbol{\check{\xi}}$. The MIMO system model can then be rewritten as
\begin{align}
&\boldsymbol{\check{y}}=\boldsymbol{A}\boldsymbol{\check{\xi}}+\boldsymbol{\check{\eta}}=\boldsymbol{\check{z}}+\boldsymbol{\check{\eta}},\label{eq:compact_general_vectorization_unquantized_MIMO_sc_data_equalization_model_block}\\
&\boldsymbol{\check{r}}=\Quant\left(\boldsymbol{\check{y}}\right),\label{eq:compact_general_vectorization_quantized_MIMO_sc_data_equalization_model_block}
\end{align}
where $\boldsymbol{A}\in\mathbb{C}^{M\cdot N_b \times P}$, $\boldsymbol{\check{\xi}}\in\mathbb{C}^{P}$, $\boldsymbol{\check{z}}, \boldsymbol{\check{y}}, \boldsymbol{\check{r}}, \boldsymbol{\check{\eta}}\in\mathbb{C}^{M\cdot N_b}$, $P = K \left(N_b+L\right)$ for the exact and $P = K N_b$ for the mismatched system model. The relationship between $\boldsymbol{\check{\xi}}$ and $\boldsymbol{\check{y}}$ in \eqref{eq:compact_general_vectorization_unquantized_MIMO_sc_data_equalization_model_block} can be described by the conditional PDF:
\begin{equation}
p\left(\boldsymbol{\check{y}}\middle|\boldsymbol{\check{\xi}}\right)
=\frac{1}{\left(\pi\sigma_{\eta}^2\right)^{M\cdot N_b}}\exp\left(-\frac{\left\|\boldsymbol{\check{y}}-\boldsymbol{A}\boldsymbol{\check{\xi}}\right\|_2^2}{\sigma_{\eta}^2}\right)
\label{eq:p_y_given_h}
\end{equation}
 as $\left.\boldsymbol{\check{y}}\middle|\boldsymbol{\check{\xi}}\right.\sim\mathcal{CN}\left(\boldsymbol{A}\boldsymbol{\check{\xi}},\sigma_\eta^2\boldsymbol{I}_{M\cdot N_b}\right)$ \cite{Kay1993}. Similarly, \eqref{eq:compact_general_vectorization_quantized_MIMO_sc_data_equalization_model_block} can be represented as the conditional \textit{Probability Mass Function} (PMF)
\begin{equation}
p\left(\boldsymbol{\check{r}}\middle|\boldsymbol{\check{y}}\right)
=\mathbb{I}_{D\left(\boldsymbol{\check{r}}\right)}\left(\boldsymbol{\check{y}}\right)
=p\left(\boldsymbol{\check{r}}\middle|\boldsymbol{\check{y}},\boldsymbol{\xi}\right)
\label{eq:p_r_given_y}
\end{equation}
of $\boldsymbol{\check{r}}$ given $\boldsymbol{\check{y}}$, where
\begin{equation}
\mathbb{I}_{D\left(\boldsymbol{\check{r}}\right)}\left(\boldsymbol{\check{y}}\right)
= \begin{cases}1, & \boldsymbol{\check{r}}=\Quant\left(\boldsymbol{\check{y}}\right) \\  0, & \text{otherwise}\end{cases}.
\label{eq:indicator_function}
\end{equation}
With (\ref{eq:p_y_given_h}) and (\ref{eq:p_r_given_y}), the joint PDF of $\boldsymbol{\check{r}}$, $\boldsymbol{\check{y}}$ and $\boldsymbol{\xi}$ reads as
\begin{equation}
p\left(\boldsymbol{\check{r}},\boldsymbol{\check{y}},\boldsymbol{\check{\xi}}\right)
\!=\frac{\mathbb{I}_{D\left(\boldsymbol{\check{r}}\right)}\left(\boldsymbol{\check{y}}\right)}{\left(\pi\sigma_{\eta}^2\right)^{M\cdot N_b}}\exp\left(-\frac{\left\|\boldsymbol{\check{y}}-\boldsymbol{A}\boldsymbol{\check{\xi}}\right\|_2^2}{\sigma_{\eta}^2}\right)
p\left(\boldsymbol{\check{\xi}}\right).
\label{eq:p_r_y_h}
\end{equation}

%% file: EM_Nonlinear_Data_Equalization.tex
\section{Expectation-Maximization (EM) based Maximum a Posteriori (MAP) Solution}
\label{sec4}
The direct maximization of $p\left(\boldsymbol{\check{r}},\boldsymbol{\check{\xi}}\right)$ using the MAP estimate
\begin{equation}
\widehat{\boldsymbol{\check{\xi}}}
= \argmax_{\boldsymbol{\check{\xi}}\in\mathbb{C}^{P}} \enspace \ln\left(p\left(\boldsymbol{\check{\xi}}\middle|\boldsymbol{\check{r}}\right)\right)
= \argmax_{\boldsymbol{\check{\xi}}\in\mathbb{C}^{P}} \enspace p\left(\boldsymbol{\check{r}},\boldsymbol{\check{\xi}}\right)
\label{eq:MAP_estimate}
\end{equation}
is in general intractable \cite{Kay1993}. The EM-algorithm computes the MAP estimate $\widehat{\boldsymbol{\check{\xi}}}$ by iteratively maximizing the MAP log-likelihood function $\ln\left(p\left(\boldsymbol{\check{r}},\boldsymbol{\check{\xi}}\right)\right)$ \cite{Kay1993,Mezghani2010}.

\subsection{EM-Algorithm} At each iteration, the following two steps are performed:
\vspace{0.2cm}
\subsubsection{Expectation (E)-step}
In the $u^{\text{th}}$ iteration of the E-step, the expected MAP log-likelihood function is computed
\begin{equation}
q\left(\boldsymbol{\check{\xi}},\widehat{\boldsymbol{\check{\xi}}}^{\left(u\right)}\right)
=\E_{\left.\boldsymbol{\check{y}}\middle|\boldsymbol{\check{r}},\widehat{\boldsymbol{\check{\xi}}}^{\left(u\right)}\right.}\left[\ln \left(p\left(\boldsymbol{\check{r}},\boldsymbol{\check{y}},\boldsymbol{\check{\xi}}\right)\right)\right].
\label{eq:q_h_h_hat_l}
\end{equation}
It is shown in \cite[c.f. Eq. (21)]{Stoeckle_EM} that \eqref{eq:q_h_h_hat_l} reduces to
\begin{equation}
\widehat{\boldsymbol{\check{y}}}^{\left(u\right)}
=\E\left[\boldsymbol{\check{y}}\middle|\boldsymbol{\check{r}},\widehat{\boldsymbol{\check{\xi}}}^{\left(u\right)}\right]
=\E_{\left.\boldsymbol{\check{y}}\middle|\boldsymbol{\check{r}},\widehat{\boldsymbol{\check{\xi}}}^{\left(u\right)}\right.}\left[\boldsymbol{\check{y}}\right],
\label{eq:e_step_complex}
\end{equation}
which is an estimate of the unquantized receive signal $\boldsymbol{\check{y}}$. A closed form expression for this expectation is derived in Appendix~\ref{app:e_and_m_step}. The $i^{\text{th}}$ element of $\widehat{\boldsymbol{\check{y}}}^{\left(u\right)}$ $i\in\left\{1,2,\ldots,M\cdot N_b \right\}$, is given by:
\begin{align}
\hat{y}_i^{(u)}=\frac{\sigma_{\eta}}{\sqrt{2}}
\left(\frac{\Re\left\{r_i\right\}\varphi\left(w_R\right)}{\normcdf\left(w_R\right)}+\textrm{j}\frac{\Im\left\{r_i\right\}\varphi\left(w_I\right)}{\normcdf\left(w_I\right)}\right) %
+ z_i,
\label{eq:e_step_elementwise_complex}
\end{align}
where $z_i=\boldsymbol{a}_i^{\T}\widehat{\boldsymbol{\check{\xi}}}^{\left(u\right)}$, $w_R=\Re\left\{r_i\right\}\Re\left\{z_i\right\}/\sqrt{\sigma_{\eta}^2/2}$, $w_I=\Im\left\{r_i\right\}\Im\left\{z_i\right\}/\sqrt{\sigma_{\eta}^2/2}$ and $r_i$ is the $i^{\text{th}}$ element of $\boldsymbol{\check{r}}$, $\boldsymbol{a}_i^{\T}$ is the $i^{\text{th}}$ row of $\boldsymbol{A}$.

\vspace{0.2cm}
\subsubsection{Maximization (M)-step}
The maximization of the expected MAP log-likelihood function $q\left(\boldsymbol{\check{\xi}},\widehat{\boldsymbol{\check{\xi}}}^{\left(u\right)}\right)$ with respect to $\boldsymbol{\check{\xi}}$ in the $u^{\text{th}}$ iteration is given as \cite[c.f. Eq. (22)]{Stoeckle_EM}:
\begin{align}
\widehat{\boldsymbol{\check{\xi}}}^{\left(u+1\right)}&=\argmin_{\boldsymbol{\check{\xi}}\in\mathbb{C}^{P}}
\left\|\boldsymbol{A}\boldsymbol{\check{\xi}}-\widehat{\boldsymbol{\check{y}}}^{\left(u\right)}\right\|_2^2
-\sigma_{\eta}^2\ln \left(p\left(\boldsymbol{\check{\xi}}\right)\right) \nonumber \\
&\overset{(a)}=\left(\boldsymbol{A}^{\h}\boldsymbol{A} + \sigma_\eta^2 \boldsymbol{R}_{\check{\boldsymbol{\xi}}\check{\boldsymbol{\xi}}}^{-1}\right)^{-1}\boldsymbol{A}^{\h}\widehat{\boldsymbol{\check{y}}}^{\left(u\right)}=\boldsymbol{G}\widehat{\boldsymbol{\check{y}}}^{\left(u\right)}.
\label{eq:m_step_complex}
\end{align}
A priori information about the vector $\boldsymbol{\check{\xi}}$ can be incorporated with the prior PDF $p\left(\boldsymbol{\check{\xi}}\right)$. A Gaussian prior is assumed in $(a)$ and the matrix $\boldsymbol{G}$ represents the space-time linear equalizer. The EM-algorithm is summarized in Algorithm~\ref{algo:EM3}.
\begin{algorithm}
\caption{Expectation-Maximization (EM) Algorithm}
\label{algo:EM3}
\begin{algorithmic}
    \REQUIRE $\boldsymbol{A}$, $\boldsymbol{\check{r}}$, $\widehat{\boldsymbol{\check{\xi}}}^{\left(0\right)}$, $\sigma_{\eta}^2$, $p\left(\boldsymbol{\check{\xi}}\right)$
		\STATE \textbf{Initialize:} $u = 0$
    \WHILE{stopping criterion not met}
        \STATE \textbf{E-step:} $\widehat{\boldsymbol{\check{y}}}^{\left(u\right)}
=\E\left[\boldsymbol{\check{y}}\middle|\boldsymbol{\check{r}},\widehat{\boldsymbol{\check{\xi}}}^{\left(u\right)}\right]
$
        \STATE \textbf{M-step:}  $\displaystyle\widehat{\boldsymbol{\check{\xi}}}^{\left(u+1\right)}\!=\!\left(\boldsymbol{A}^{\h}\boldsymbol{A}+\sigma_\eta^2 \boldsymbol{R}_{\check{\boldsymbol{\xi}}\check{\boldsymbol{\xi}}}^{-1} \right)^{-1}\boldsymbol{A}^{\h}\widehat{\boldsymbol{\check{y}}}^{\left(u\right)}$
				\STATE $u := u+1$
    \ENDWHILE
    \ENSURE $\widehat{\boldsymbol{\check{\xi}}}^{\left(u\right)}$
\end{algorithmic}
\end{algorithm}

The EM-algorithm can be stopped after $I_{\text{max}}$ iterations or in the case $\left\|\widehat{\boldsymbol{\check{\xi}}}^{\left(u\right)}-\widehat{\boldsymbol{\check{\xi}}}^{\left(u-1\right)}\right\|_2 \leq \gamma_{\text{EM}}{\left\|\widehat{\boldsymbol{\check{\xi}}}^{\left(u\right)}\right\|_2}$ with $\gamma_{\text{EM}} > 0$.

\subsection{Computational Complexity}

The necessary number of complex multiplications in relation to the coherence time $T_c$ is derived in this section. It is assumed that the coherence time represented by $T_{c}$ symbols is divided into $B$ blocks, each consisting of $N_b$ symbols:
\begin{equation}
T_c=B\cdot N_b. \label{eq:coherence time definition}
\end{equation}
In the following we distinguish between static and dynamic complexity. As $\boldsymbol{G}$ in \eqref{eq:m_step_complex} needs to be computed only \emph{once} during $T_c$,  its computational complexity is
\begin{equation}
\mathcal{T}_{\textrm{G}}=P^3+P^2\cdot M \cdot N_b. \label{eq:computational_complexity_G}
\end{equation}
The dynamic complexity is a matrix-vector product calculated from the E-step \eqref{eq:e_step_complex} and M-step \eqref{eq:m_step_complex} for the $v^{\textrm{th}}$ block as
\begin{equation}
\mathcal{T}^{(v)}_{\textrm{E}}+\mathcal{T}^{(v)}_{\textrm{M}}\!=M\cdot N_b\cdot \left(P+1\right)+P\cdot M \cdot N_b, \label{eq:computational_complexity_E_M_step}
\end{equation}
where $v\in\left\{1,\ldots,B\right\}$. The computational complexity of the Gaussian PDF $\varphi(\boldsymbol{\cdot})$ and \textit{Cumulative Distribution Function} (CDF) $\normcdf(\boldsymbol{\cdot})$ are ignored in \eqref{eq:computational_complexity_E_M_step} as the latter can be pre-calculated offline. With \eqref{eq:computational_complexity_G} and \eqref{eq:computational_complexity_E_M_step}, the total complexity of the EM-algorithm during $T_c$ is given as:
\begin{align}
\mathcal{T}_{\textrm{tot}}&=\sum_{v=1}^{B} I_{v} \cdot \left[\mathcal{T}^{(v)}_{\textrm{E}}+\mathcal{T}^{(v)}_{\textrm{M}}\right] + \mathcal{T}_{\textrm{G}}\nonumber\\
													&=\sum_{v=1}^{B} I_{v} \left[M\cdot N_b\cdot \left(2P+1\right)\right]+P^3+P^2\cdot M \cdot N_b, \label{eq:tot_computational_complexity_EM_algorithm}
\end{align}
where $I_{v}$ is the number EM-iterations until convergence for the $v$-block.
As $P \propto N_b$ for both models, furthermore assuming a constant $I_{v}$, i.e. $I_{v} \approx I \; \forall\; v$, the computational complexity for the static and dynamic part are $\mathcal{O}\left( N_b^3 \left( K^3 + K^2 M \right)\right)$ and $\mathcal{O}\left(2N_b^2 B I K M \right)$, respectively. It is shown in \cite{Munir_Plabst18} that $N_b \propto L$.  Therefore,  running the EM-algorithm in time-domain using \eqref{eq:e_step_complex} and \eqref{eq:m_step_complex} becomes computationally infeasible for  $N_b\!\gg\!L$.

%% file: Computational_Complexity.tex
\section{The EM-Algorithm in Frequency Domain }\label{sec5}
This section deals with the reduction of computational complexity of the EM-algorithm by exploiting the block-circulant structure of the channel matrix $\boldsymbol{A}=\boldsymbol{\check{H}}_{\textrm{cir}}$ using the mismatched model in \eqref{eq:e_step_complex} and \eqref{eq:m_step_complex}. Therefore, $\boldsymbol{\check{H}}_{\textrm{cir}}$ can be diagonalized by the \textit{Discrete Fourier Transform} (DFT),
\begin{equation}
\boldsymbol{\check{H}}_{\textrm{cir}}=\left(\boldsymbol{F}^{\h}\otimes \boldsymbol{I}_M\right)\boldsymbol{\mathcal{H}}\left(\boldsymbol{F}\otimes \boldsymbol{I}_K\right)\label{eq:block_circulant},
\end{equation}
where $\boldsymbol{F}\in\mathbb{C}^{N_b\times N_b}$ is an $N_b$-point DFT-matrix and $\boldsymbol{\mathcal{H}}$ is block-diagonal:
\begin{equation}
\boldsymbol{\mathcal{H}}=\textrm{diag}{\left\{\boldsymbol{H}_{f_i}\right\}}_{i=1}^{N_b}, \textrm{ where} \label{eq:frequency_complete_channel_matrix}
\end{equation}
\begin{equation*}
\boldsymbol{H}_{f_i}\!=\!\sum_{l=0}^{L}\boldsymbol{H}_l\cdot \textrm{exp}\left(-\textrm{j}\!\cdot \frac{2\pi}{N_b}l(i-1)\right)\!, \textrm{ for } 1\!\leq\!i\!\leq\!N_b,
\end{equation*}
represents the multi-path MIMO channel in frequency domain.
\subsection{Efficient E-Step} Applying \eqref{eq:block_circulant} in $\boldsymbol{\check{z}}=\boldsymbol{\check{H}}_{\textrm{cir}}\boldsymbol{\check{\xi}}$ (c.f. Eq. \eqref{eq:compact_general_vectorization_unquantized_MIMO_sc_data_equalization_model_block}), the vector $\boldsymbol{\check{z}}$ is calculated in frequency domain
\begin{equation}
\boldsymbol{\check{z}}=\left(\boldsymbol{F}^{\h}\otimes \boldsymbol{I}_M\right)\boldsymbol{\mathcal{H}}\left(\boldsymbol{F}\otimes \boldsymbol{I}_K\right)\boldsymbol{\check{\xi}}\label{eq:noiseless_unquantized_z}.
\end{equation}

\subsection{Efficient M-Step} Applying \eqref{eq:block_circulant} in \eqref{eq:m_step_complex}, the matrix $\boldsymbol{G}$ can be efficiently calculated in frequency domain:
\begin{align}
\boldsymbol{G}=\left(\boldsymbol{F}^{\h}\otimes \boldsymbol{I}_K\right)\boldsymbol{G}_f\left(\boldsymbol{F}\otimes \boldsymbol{I}_M\right)\label{eq:G_equalizer_using_Gf},
\end{align}
where $\boldsymbol{G}_f$ represents the frequency domain equalizer:
\begin{align}
\boldsymbol{G}_f &=\left(\boldsymbol{\mathcal{H}}^{\h}\boldsymbol{\mathcal{H}} + \frac{\sigma_\eta^2}{\sigma_x^2} \boldsymbol{I}_{K N_b} \right)^{-1}\!\boldsymbol{\mathcal{H}}^{\h}\\
&= \textrm{diag}{\left\{\left(\boldsymbol{H}_{f_i}^{\h}\boldsymbol{H}_{f_i} + \frac{\sigma_\eta^2}{\sigma_x^2} \boldsymbol{I}_K \right)^{-1}\!\boldsymbol{H}_{f_i}^{\h}\right\}}_{i=1}^{N_b}.\label{eq:frequency_domain_equalizer}
\end{align}
With \eqref{eq:G_equalizer_using_Gf} in \eqref{eq:m_step_complex}, the efficient M-step can be implemented as:
\begin{itemize}
\item Frequency domain conversion of $\widehat{\boldsymbol{\check{y}}}^{\left(u\right)}$:
\begin{align}
\widehat{\boldsymbol{\check{y}}_f}^{\left(u\right)}=\left(\boldsymbol{F}\otimes \boldsymbol{I}_M\right)\widehat{\boldsymbol{\check{y}}}^{\left(u\right)}\label{eq:fft_ycheck}.
\end{align}
\item Frequency domain equalization:
\begin{align}
\widehat{\boldsymbol{\check{\xi}}_f}^{\left(u+1\right)}=\boldsymbol{G}_f\widehat{\boldsymbol{\check{y}}_f}^{\left(u\right)}\label{eq:freq_dom_equal}.
\end{align}
\item Time domain conversion of $\widehat{\boldsymbol{\check{\xi}}_f}^{\left(u+1\right)}$
\begin{align}
\widehat{\boldsymbol{\check{\xi}}}^{\left(u+1\right)}=\left(\boldsymbol{F}^{\h} \otimes \boldsymbol{I}_K\right)\widehat{\boldsymbol{\check{\xi}}_f}^{\left(u+1\right)}\label{eq:ifft_xichek}.
\end{align}
\end{itemize}
Note, that the term $\boldsymbol{G}\boldsymbol{\check{\gamma}}'[n]$ in \eqref{eq:vectorization_unquantized_MIMO_sc_data_equalization_model_block_circulant} is an interference distortion, which corrupts the whole estimated data-block $\widehat{\boldsymbol{\check{\xi}}}$. The next subsection deals with minimizing the interference distortion.
\subsection{Interference Analysis}\label{sec:interf_analysis}
It was shown in \cite{Hueske_2009} that the ensemble-averaged \emph{interference distortion} power has a bathtub-like distribution. This behavior can be exploited to minimize the resulting error by using a $L'$ samples overlapping of data blocks, i.e., $\boldsymbol{R}[n]$ contains vectors corresponding to the time instances $n,\ldots,n-(N_b-1)$ and is followed by $\boldsymbol{R}[n\!-(N_b\!-L')]$ with corresponding time elements $n\!-(N_b\!-L'),\ldots,n\!-(2N_b\!-L'-1)$.
In this work we will show that $L'$  is directly related to the length of the channel memory  \cite{Leibrich2010}, i.e., $L' \propto L$.  Since $L'$ can be even or odd, we define the pre-discard- and post-discard-lengths as $L_\text{pre} = \lceil L'/2 \rceil$ and $L_\text{post} = \lfloor L'/2 \rfloor$.
\subsection{Computational Complexity}
Let $\mathcal{T}^{(v)}_{\textrm{E}_f}$, $\mathcal{T}^{(v)}_{\textrm{M}_f}$ and $\mathcal{T}_{\textrm{G}_f}$ denote the computational complexity of the E-step, M-step and calculation of the matrix $\boldsymbol{G}$ in frequency domain. With a $L'$ samples overlapping of data-blocks, we will process $B' = T_c/\left(N_b - L' \right)$ blocks per $T_c$.
Using \Cref{eq:frequency_domain_equalizer,eq:fft_ycheck,eq:freq_dom_equal,eq:ifft_xichek}, the total complexity during $T_c$ for performing FDE with the EM-algorithm evaluates to:
\begin{align}
\mathcal{T}_{\textrm{tot}_{f}}&= \sum_{v=1}^{B'} I_v \cdot \left[\mathcal{T}^{(v)}_{\textrm{E}_f}+\mathcal{T}^{(v)}_{\textrm{M}_f}\right] + \mathcal{T}_{\textrm{G}_f}\nonumber\\
													&=\sum_{v=1}^{B'}I_v \bigg[\left(\left(M + K\right)\cdot {\log_2{N_b}}+ K\cdot M\right) \cdot N_b \bigg] \cdot 2 \nonumber \\
													&\phantom{ab}+ \left( KM \log_2\left(N_b\right) + 2\cdot K^2M + K^3  \right) \cdot N_b. \label{eq:tot_computational_complexity_frequency_domain_EM_algorithm}
\end{align}
For derivation of $\mathcal{T}_{\textrm{G}_f}$, we refer to \cite{Munir_Plabst18}, Eq. (31), with the slight modification that the complexity regarding the noise-covariance matrix is not accounted for in \eqref{eq:tot_computational_complexity_frequency_domain_EM_algorithm}, as it is a scaled identity matrix \cite{Stoeckle_EM}.
The terms $\mathcal{T}^{(v)}_{\textrm{E}_f}$ and $\mathcal{T}^{(v)}_{\textrm{M}_f}$ are equal, and describe the FFT and IFFT of the data-blocks, both needed in the E-step and the M-step.

Therefore, the static and dynamic computational complexity become log-linear in $N_b$, effectively reducing the computational complexity by orders of magnitude compared to \eqref{eq:tot_computational_complexity_EM_algorithm}.

%% file: Simulation_Results_And_Analysis.tex
\section{Simulation Results and Analysis}\label{sec6}
Consider a MIMO setup having $K$ single transmit antenna users and $M$ receive antennas. The CIR between each pair of transmit and receive antenna consists of $L\!+\!1$ taps.
The transmitter is employing 16-QAM, CP is omitted at the transmit side and the receiver is using $1$-bit quantizers at each of the receive antennas. The channel coherence time is assumed to be $T_c=\SI{50e3}{}$ symbols. The noise is i.i.d. zero-mean circularly-symmetric AWGN with variance $\sigma_{\eta}^2=1$ (cf. Sec. \ref{sec2}) and the channel is chosen based on an Extended Vehicular A model ($9$ nonzero taps). The results are averaged over $N_\text{sim} = 200$ channel realizations. Perfect CSI and synchronization between transmitter and receiver is assumed throughout the equalization process. For the EM-algorithm, the stopping criterion is set to $\gamma_\text{EM} = \SI{1e-3}{}$ and the maximum number of iterations is upper bounded by $I_\text{max} = 1000$.

We define the bit energy to noise spectral density $\!E_b\!/\!N_0\!$ as:
\begin{equation}
\frac{E_b}{N_0}=\frac{P_t}{KM\sigma_{\eta}^2}\frac{\textrm{trace}\left(\E_{\left.\boldsymbol{\check{H}}\right.}\left(\boldsymbol{\check{H}}\boldsymbol{\check{H}}^H\right)\right)}{B},
\label{eq:sim_EbNo}
\end{equation}
where $P_t$ is the total transmit power, $B$ is the number of bits per constellation symbol and $\boldsymbol{\check{H}}$ is the MIMO channel matrix.

The proposed EM-algorithm with different settings of overlap-discard processing is assumed throughout this section. A comparison to linear equalization based on \cite{Munir_Plabst18} is established for performance comparison. The linear WF-approach for quantized and unquantized MIMO systems from \cite{Munir_Plabst18} is denoted as WF$_\text{E,Q}$ and WF$_\text{E}$ respectively, whereas our nonlinear EM-approach for the exact and mismatched model is denoted as EM$_{\mu}\; \text{for } \mu \in \lbrace \text{E,M}\rbrace$ and initialized with WF$_\text{$\mu$,Q},\, \mu \in \lbrace \text{E,M}\rbrace$ respectively, if not otherwise noted.
\subsubsection{Comparison between linear and nonlinear equalization methods}
\begin{figure}\centering
\input{includes/Figures/BER_vs_Discarding_Factor.tex}
\caption{BER as a function of $E_b/N_0$, different discarding factors $L'\!=\!2L,\,\!3L\!$ and equalization block lengths $N_b\!=\!512$ (dashed), $1024$ (dotted). $K\!=\!2$, $M\!=\!32$, CIR $L\!+\!1\!=\!128$.}
\label{fig:BER_Different_Discarding_Factors_And_Block_Lengths}
\end{figure}
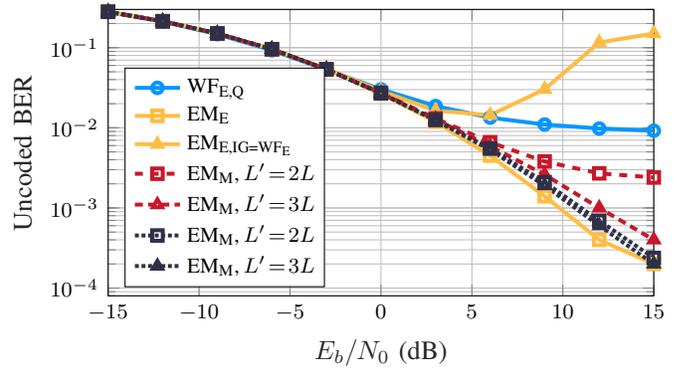

Fig.~\ref{fig:BER_Different_Discarding_Factors_And_Block_Lengths} depicts the BER of an EM-based FDE block-processor for different equalization block lengths $N_b$, discarding factors $L'$ and \textit{Initial Guesses} (IG). We compare our results to the linear Wiener-Filter WF$_\text{E,Q}$ from \cite{Munir_Plabst18}.

The BER-performance improves with increasing $N_b$ and the number of discarded samples $L'$. Performing EM-FDE with $N_b = 1024$ and $L' = 3L$ achieves almost the same BER performance as choosing  $N_b = T_c$, which is running the exact EM-algorithm $\text{EM}_\text{E}$. The convergence of the EM-algorithm is very sensitive to a good initial guess. On the one hand, Fig.~\ref{fig:BER_Different_Discarding_Factors_And_Block_Lengths} shows that the performance of the EM-algorithm degrades if $\text{WF}_\text{E}$ is taken as an IG. On the other hand, a substantial improvement in performance compared to linear equalization \cite{Munir_Plabst18} is achieved, if the algorithm is initialized with $\text{WF}_\text{E,Q}$.

\subsubsection{Fixed number of EM iterations}
The convergence of the EM algorithm depends strongly on the number of iterations.

\begin{figure}\centering
\input{includes/Figures/BER_vs_Maxiter.tex}
\caption{BER using different EM models as a function of $E_b/N_0$ and fixed number of EM-iterations $I_\text{max}$. $K\!=\!2$, $M\!=\!32$, CIR  $L\!+\!1\!=\!128$, $L'\!=\!2L$ and $N_b\!=\!1024$ for $\text{EM}_\text{M}$.}
\label{fig:BER_fixed_maxiter_comparison}
\end{figure}
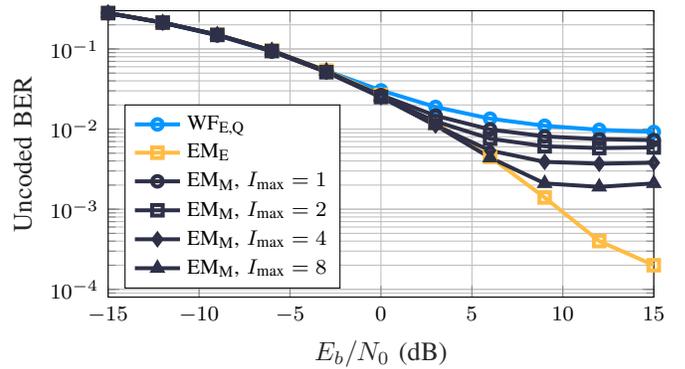

Assuming that only a limited processing-power budget is available at the base-station, we investigate running the EM-algorithm for a fixed number of iterations.
Fig.~\ref{fig:BER_fixed_maxiter_comparison} shows that the performance of $\text{EM}_\text{M}$  improves after each iteration, taking $\text{WF}_\text{M,Q}$ as an initial guess. Moreover, the performance of $\text{EM}_\text{M}$ comes quite close to $\text{EM}_\text{E}$ in the mid-SNR range after $8$ iterations. Note that the performance of $\text{EM}_\text{E}$ is taken as a benchmark, as it takes $N_b=T_c$ and $I_\text{max}=1000$. The results indicate that the proposed $\text{EM}_\text{M}$  achieves the same performance as $\text{EM}_\text{E}$ with substantially reduced complexity.

%% file: includes/Figures/BER_vs_Discarding_Factor.tex
\definecolor{b}{HTML}{0496ff}%
\definecolor{o}{HTML}{2d3047}%
\definecolor{pur}{HTML}{ffbc42}%

\begin{tikzpicture}

\begin{axis}[%
  width=0.40\textwidth,
  height=1.5in,
at={(1.139in,0.544in)},
scale only axis,
xmin=-15,
xmax=15,
xlabel style={font=\color{white!15!black}},
xlabel={$E_b/N_0 \text{ (dB)}$},
ymode=log,
ymin=8E-5,
ymax=0.3,
yminorticks=true,
ylabel style={font=\color{white!15!black}},
ylabel={Uncoded BER},
axis background/.style={fill=white},
title style={font=\bfseries},
xmajorgrids,
ymajorgrids,
yminorgrids,
legend cell align=left,
legend pos=south west,
]

\addplot [color=b, mark=o, mark options={solid,b} ,line width=1.5pt, solid]
  table[row sep=crcr]{%
-15	 0.2790  \\
-12	0.2112\\
-9	0.1472\\
-6	 0.0927 \\
-3	0.0532\\
0	 0.0302\\
3 0.0187\\
6	 0.0134\\
9	 0.0110 \\
12	 0.0098 \\
15	0.0092 \\
};
\addlegendentry{$\text{WF}_\text{E,Q}$}

\addplot [color=pur, line width=1.5pt, mark=square, mark options={solid}, solid]
  table[row sep=crcr]{%
-15 0.2805\\
-12	0.2136\\
-9	0.1505\\
-6	0.0957\\
-3	0.0537\\
0	 0.0267\\
3	 0.0119\\
6	 0.0045\\
9	 0.0014\\
12	0.0004\\
15 0.0002\\
};
\addlegendentry{$\text{EM}_\text{E}$\phantom{$\,$}}

\addplot [color=pur, line width=1.5pt, mark=triangle, mark options={solid}, solid]
  table[row sep=crcr]{%
-15 0.2805\\
-12	0.2136\\
-9	0.1507\\
-6	0.0961\\
-3	0.0546\\
0	 0.0288\\
3	  0.0164\\
6	  0.0144\\
9	  0.0303\\
12	0.1162 \\
15  0.1502\\
};
\addlegendentry{$\text{EM}_\text{E,IG=WF$_\text{E}$}$}
\addplot [color=ppink, dashed, line width=1.5pt, mark=square, mark options={solid}]
  table[row sep=crcr]{%
-15	0.2805 \\
-12	 0.2136\\
-9	0.1505\\
-6	0.0958\\
-3	0.0538\\
0	 0.0272\\
3	 0.0131 \\
6  0.0066\\
9	 0.0038 \\
12	0.0027 \\
15 0.0024	 \\
};
\addlegendentry{$\text{EM}_\text{M}$,$\,L'\!=\! 2L\!$}

\addplot [color=ppink, densely dashed, line width=1.5pt, mark=triangle, mark options={solid}]
  table[row sep=crcr]{%
-15	0.2805 \\
-12	0.2136\\
-9	0.1505\\
-6	0.0958\\
-3	0.0538 \\
0	0.0270\\
3	0.0127 \\
6	 0.0059\\
9	0.0026\\
12	 0.0010\\
15	0.0004 \\
};
\addlegendentry{$\text{EM}_\text{M}$,$\,L'\!=\!3L\!$}

\addplot [color=o, densely dotted, line width=1.5pt, mark=square, mark options={solid}]
  table[row sep=crcr]{%
-15	0.2805\\
-12	 0.2136\\
-9	0.1505\\
-6	0.0958\\
-3	0.0538\\
0	0.0270\\
3	  0.0125\\
6	 0.0055\\
9	0.0021 \\
12	0.6927E-3 \\
15	 0.2376e-03\\
};
\addlegendentry{$\text{EM}_\text{M}$,$\,L'\!=\!2L\!$}
\addplot [color=o, densely dotted, line width=1.5pt, mark=triangle, mark options={solid}]
  table[row sep=crcr]{%
-15	 0.2805\\
-12	0.2136 \\
-9	0.1505\\
-6	 0.0958\\
-3	0.0537\\
0	 0.0269\\
3	 0.0124\\
6	 0.0053\\
9	0.0019\\
12	0.0006 \\
15	 0.0002\\
};
\addlegendentry{$\text{EM}_\text{M}$,$\,L'\!=\!3L\!$}

\end{axis}
\end{tikzpicture}%

%% file: includes/Figures/BER_vs_Maxiter.tex
\definecolor{r}{HTML}{c0392b}%
\definecolor{b}{HTML}{0496ff}%
\definecolor{bblack}{HTML}{2d3047}%
\definecolor{o}{HTML}{ffbc42}%

\begin{tikzpicture}

\begin{axis}[%
  width=0.40\textwidth,
  height=1.5in,
at={(1.139in,0.544in)},
scale only axis,
xmin=-15,
xmax=15,
xlabel style={font=\color{white!15!black}},
xlabel={$E_b/N_0 \text{ (dB)}$},
ymode=log,
ymin=8E-5,
ymax=0.30,
yminorticks=true,
ylabel style={font=\color{white!15!black}},
ylabel={Uncoded BER},
axis background/.style={fill=white},
title style={font=\bfseries},
xmajorgrids,
ymajorgrids,
yminorgrids,
legend cell align=left,
legend pos=south  west,
]

\addplot [color=b, mark=o, mark options={solid} ,line width=1.5pt]
  table[row sep=crcr]{%
-15	0.2795 \\
-12	0.2116\\
-9	0.1476\\
-6	 0.0931 \\
-3	0.0535\\
0	0.0304\\
3	0.0189\\
6	0.0135\\
9	0.0110 \\
12	0.0098\\
15	0.0092\\
};
\addlegendentry{$\text{WF}_\text{E,Q}$\phantom{$\,$}}

\addplot [color=o, line width=1.5pt, mark=square, mark options={solid}]
  table[row sep=crcr]{%
-15 0.2805\\
-12	0.2136\\
-9	0.1505\\
-6	0.0957\\
-3	0.0537\\
0	 0.0267\\
3	 0.0119\\
6	 0.0045\\
9	 0.0014\\
12	0.0004\\
15 0.0002\\
};
\addlegendentry{$\text{EM}_\text{E} $}

\addplot [color=bblack, solid, line width=1.5pt, mark=o, mark options={solid}]
  table[row sep=crcr]{%
-15	 0.2802\\
-12	0.2128\\
-9	0.1489  \\
-6  0.0933\\
-3	0.0515\\
0	 0.0266 \\
3	 0.0147\\
6	 0.0099\\
9	 0.0081\\
12 0.0075\\
15 0.0073\\
};

\addlegendentry{$\text{EM}_\text{M}$, $I_\text{max} = 1$}

\addplot [color=bblack, solid, line width=1.5pt, mark=square, mark options={solid}]
  table[row sep=crcr]{%
-15	0.2804 \\
-12	0.2133\\
-9	0.1497 \\
-6	0.0941\\
-3	0.0513 \\
0	 0.0251\\
3	  0.0125  \\
6	0.0076\\
9	0.0061\\
12 0.0058\\
15 0.0059\\
};
\addlegendentry{$\text{EM}_\text{M}$, $I_\text{max}  = 2$}

\addplot [color=bblack, solid, line width=1.5pt, mark=diamond, mark options={solid}]
  table[row sep=crcr]{%
-15	0.2805\\
-12	 0.2135\\
-9	 0.1504\\
-6	0.0951  \\
-3	 0.0521   \\
0	 0.0250\\
3	0.0111\\
6	0.0054\\
9	0.0039\\
12 0.0037\\
15   0.0038\\
};
\addlegendentry{$\text{EM}_\text{M}$, $I_\text{max} = 4$}

\addplot [color=bblack, solid, line width=1.5pt, mark=triangle, mark options={solid}]
  table[row sep=crcr]{%
-15	0.2805\\
-12	 0.2135\\
-9	 0.1504\\
-6	0.0957 \\
-3	 0.0533  \\
0	 0.0255\\
3	 0.0111\\
6	 0.0044\\
9	 0.0021\\
12 0.0019\\
15  0.0021\\
};
\addlegendentry{$\text{EM}_\text{M}$, $I_\text{max} = 8$}

\end{axis}
\end{tikzpicture}%

%% file: Conclusion.tex
\section{Conclusion}\label{sec7}

This paper applies data detection using the EM-algorithm in $1$-bit quantized massive MIMO systems without CP. We propose a computationally efficient hybrid time-frequency approach where the E-step is performed in the time domain and the M-step in the frequency domain using the mismatched model. The interference distortion due to the block-circulant channel matrix approximation is minimized by selecting block-length $N_b$ and overlapping factor $L'$ in relation to the CIR length $L$. Numerical results show that $N\approx 4\cdot L$ and $L'\approx 2\cdot L$  is a good choice.

The simulation results show that the initial guess of the estimated data is crucial for convergence of the EM-algorithm. It is shown that taking the WF-estimate for quantized systems as an initial guess is a better choice compared to taking the WF-estimate for unquantized systems. The results also indicate that the performance of the EM-algorithm improves after each iteration under the condition that $N_b$, $L'$ and the initial guess are chosen optimally.

Future work might include the investigation of the proposed methods under non-perfect CSI. 

%% file: Derivation_of_E_and_M_Step.tex
\section{Derivation of E-step}\label{app:e_and_m_step}
According to (\ref{eq:compact_general_vectorization_unquantized_MIMO_sc_data_equalization_model_block}), the $i^{\text{th}}$ element of $\boldsymbol{\check{y}}$ is $y_i=\boldsymbol{a}_i^{\T}\boldsymbol{\check{\xi}}+\eta_i$, where $\eta_i\sim\mathcal{CN}\left(0,\sigma_\eta^2\right)$ is the $i^{\text{th}}$ element of $\boldsymbol{\eta}$, and, according to (\ref{eq:compact_general_vectorization_quantized_MIMO_sc_data_equalization_model_block}), $r_i=\Quant\left(y_i\right)$, $i\in\left\{1,2,\ldots,M\cdot N_b\right\}$.
Hence, the $i^{\text{th}}$ element of $\widehat{\boldsymbol{\check{y}}}^{\left(u\right)}$ in the E-step (\ref{eq:e_step_complex}),
\begin{equation}
  \nonumber
\hat{{y}}_i^{\left(u\right)}
=\E_{\left.\boldsymbol{\check{y}}\middle|\boldsymbol{\check{r}},\widehat{\boldsymbol{\check{\xi}}}^{\left(u\right)}\right.}\left[y_i\right]
=\E_{\left.y_i\middle|r_i,\widehat{\boldsymbol{\check{\xi}}}^{\left(u\right)}\right.}\left[y_i\right]
=\E\left[{y}_i\middle|r_i,\boldsymbol{a}_i^{\T}\widehat{\boldsymbol{\check{\xi}}}^{\left(u\right)}\right],
\label{eq:e_step_elementwise_complex_0}
\end{equation}
is a conditional expectation of the form $\E\left[y\middle|r,z\right]$ with
$y=z+\eta$,
$\eta\sim\mathcal{CN}\left(0,\sigma_\eta^2\right)$, and
$r=\Quant\left(y\right)$.
With $r_R=\Re\left\{r\right\}$, $r_I=\Im\left\{r\right\}$, $y_R=\Re\left\{y\right\}$, $y_I=\Im\left\{y\right\}$, $z_R=\Re\left\{z\right\}$, $z_I=\Im\left\{z\right\}$, $\eta_R=\Re\left\{\eta\right\}$ and $\eta_I=\Im\left\{\eta\right\}$, we arrive at
\begin{align}
y_l&=z_l+\eta_l, \textrm{ where } \quad \eta_l\sim\mathcal{N}\left(0,\sigma_\eta^2/2\right) \textrm{ and }
\label{eq:y_l} \\
r_l&=\sign\left(y_l\right),
\label{eq:r_l}
\end{align}
for $l\in\left\{R,I\right\}$. As a consequence,
\begin{equation}
\E\left[y\middle|r,z\right]=\E\left[y_R\middle|r_R,z_R\right]+\textrm{j}\E\left[y_I\middle|r_I,z_I\right]
\label{eq:E_y_given_r_z}
\end{equation}
with \vspace{-0.4cm}
\begin{equation}
\E\left[y_l\middle|r_l,z_l\right]
=\frac{\int\limits_{-\infty}^{+\infty}
{y}_l p\left(r_l,y_l,z_l\right)
\,\text{d}y_l}{\int\limits_{-\infty}^{+\infty}
p\left(r_l,y_l,z_l\right)
\,\text{d}y_l}.
\label{eq:e_step_elementwise_0}\vspace{-0.5cm}
\end{equation}
Since
\begin{equation}
p\left(r_l,y_l,z_l\right)=
p\left(r_l\middle|y_l\right)
p\left(y_l\middle|z_l\right)
p\left(z_l\right), \vspace{-0.1cm}
\end{equation}
where $\left.y_l \middle| z_l\right.\sim\mathcal{N}\left(z_l,\sigma_\eta^2/2\right)$ according to (\ref{eq:y_l}) such that
\begin{equation}
p\left(y_l\middle|z_l\right)
=\frac{1}{\sqrt{\sigma_{\eta}^2/2}}\varphi\left(\frac{y_l-z_l}{\sqrt{\sigma_{\eta}^2/2}}\right),
\label{eq:y_l_given_z_l}
\end{equation}
where $\varphi\left(x\right)=\frac{1}{\sqrt{2\pi}}\exp\left(-\frac{x^2}{2}\right)$. The relationship between $y_l$ and $r_l$ in (\ref{eq:r_l}) is reflected by
\begin{equation}
  \nonumber
p\left(r_l\middle|y_l\right)
=\begin{cases}1, & r_l=\sign\left(y_l\right) \\  0, & \text{otherwise}\end{cases}
=\begin{cases}1, & y_l\in\left(r_l^{\text{lo}},r_l^{\text{up}}\right] \\  0, & \text{otherwise}\end{cases}
\label{eq:r_l_given_y_l},
\end{equation}
with \vspace{-0.2cm}
\begin{equation}
r_l^{\text{lo}}=
\begin{cases}
-\infty,&r_l=-1\\
0,&r_l=+1
\end{cases}
\text{ and }
r_l^{\text{up}}=
\begin{cases}
0,&r_l=-1\\
+\infty,&r_l=+1
\end{cases}.
\label{eq:r_loup}
\end{equation}
The conditional expectation \eqref{eq:e_step_elementwise_0} can be written as \vspace{-0.2cm}
\begin{equation}
\E\left[y_l\middle|r_l,z_l\right]
=\frac{\int\limits_{r_l^{\text{lo}}}^{r_l^{\text{up}}}
{y}_l \varphi\left(\frac{y_l-z_l}{\sqrt{\sigma_{\eta}^2/2}}\right)
\,\text{d}y_l}{\int\limits_{r_l^{\text{lo}}}^{r_l^{\text{up}}}
\varphi\left(\frac{y_l-z_l}{\sqrt{\sigma_{\eta}^2/2}}\right)
\,\text{d}y_l}.
\label{eq:e_step_elementwise_1}
\end{equation}
Evaluating the integrals in (\ref{eq:e_step_elementwise_1}) gives the expression \vspace{-0.2cm}
\begin{equation}
\E\left[y_l\middle|r_l,z_l\right]=r_l\frac{\sigma_{\eta}}{\sqrt{2}}\frac{\varphi\left(\frac{r_lz_l}{\sqrt{\sigma_{\eta}^2/2}}\right)}{\normcdf\left(\frac{r_lz_l}{\sqrt{\sigma_{\eta}^2/2}}\right)}+z_l
\label{eq:e_step_elementwise_2},
\end{equation}
where $\normcdf\left(x\right)=\int_{-\infty}^{x}\varphi\left(t\right)\,\text{d}t$. The conditional expectation in (\ref{eq:E_y_given_r_z}) can be expressed as
\begin{equation}
  \nonumber
\begin{split}
&\E\left[y\middle|r,z\right]
=\frac{\sigma_{\eta}}{\sqrt{2}}\left(r_R\frac{\varphi\left(\frac{r_Rz_R}{\sqrt{\sigma_{\eta}^2/2}}\right)}{\normcdf\left(\frac{r_Rz_R}{\sqrt{\sigma_{\eta}^2/2}}\right)}
+\textrm{j}\cdot r_I\frac{\varphi\left(\frac{r_Iz_I}{\sqrt{\sigma_{\eta}^2/2}}\right)}{\normcdf\left(\frac{r_Iz_I}{\sqrt{\sigma_{\eta}^2/2}}\right)}\right)+z.
\end{split}
\end{equation}
Using this result in (\ref{eq:E_y_given_r_z}) with $y_i$, $r_i=\Re\left\{r_i\right\}+\textrm{j}\cdot \Im\left\{r_i\right\}$ and $\boldsymbol{a}_i^{\T}\widehat{{\boldsymbol{\check{\xi}}}}^{\left(u\right)}=\Re\left\{\boldsymbol{a}_i^{\T}\widehat{{\boldsymbol{\check{\xi}}}}^{\left(u\right)}\right\}+\textrm{j}\cdot \Im\left\{\boldsymbol{a}_i^{\T}\widehat{{\boldsymbol{\check{\xi}}}}^{\left(u\right)}\right\}$ instead of $y$, $r=r_R+\textrm{j}\cdot r_I$ and $z=z_R+\textrm{j}\cdot z_I$, respectively, results in the elementwise computation of the E-step (\ref{eq:e_step_elementwise_complex}).